\documentclass{article}
\usepackage{pdfsync}
\usepackage{graphicx}
\usepackage[english]{babel}
\usepackage[a4paper]{geometry}
\usepackage{amsmath,bm}
\usepackage{textcomp}
\usepackage[normalsize]{subfigure}

\def\..{\,\mathpunct{\ldotp\ldotp}} 

\newcommand{\TS}{T}

\newtheorem{theorem}{Theorem}

\newtheorem{corollary}{Corollary}

\newcounter{noqed}
\newcommand{\qed}{ \ifmmode\mbox{ }\fi\rule[-.05em]{.3em}{.7em}\setcounter{noqed}{0}}
\newenvironment{proof}[1][{}]{\noindent{\bf Proof#1. }\setcounter{noqed}{1}}{\ifnum\value{noqed}=1\qed\fi\par\medskip}

\title{$E=I+\TS$\\The internal extent formula for compacted tries}
\author{Paolo Boldi\qquad Sebastiano Vigna\\Universit\`a degli Studi di Milano,
Italy}
\date{}
\begin{document}
\bibliographystyle{alpha}

\maketitle

\begin{abstract}
It is well known~\cite[pages 399--400]{KnuACPI} that in a binary tree the
external path length minus the internal path length
is exactly $2n-2$, where $n$ is the number of external nodes. We show that a
generalization of the formula holds for compacted tries, replacing the
role of \emph{paths} with the notion of \emph{extent}, and the value $2n-2$ with the \emph{trie measure},
an estimation of the number of bits that are necessary to
describe the trie. 
\end{abstract}

\section{Introduction}

The well-known formula~\cite[pages 399--400]{KnuACPI} 
\[
	E = I + 2n - 2,
\]
where $n$ is the number of external nodes, relates the \emph{external path
length} $E$ of a binary tree (the sum of the lengths of the paths leading to external nodes) with the 
\emph{internal path length} $I$ (the sum of the lengths of the paths leading to
internal nodes).\footnote{The formula actually reported by Knuth is slightly
different ($E=I+2n$) because in his notation $n$ is the number of
\emph{internal} nodes, which is equal to the number of external nodes minus one. As we will see, for compacted tries the number of external nodes is equal to the size of the set of strings represented
by the trie, and so it is a more natural candidate for the letter $n$.} 

A \emph{compacted (binary) trie} is a binary tree where each node (both internal
and external) is endowed with a (binary) string (possibly empty) called
\emph{compacted path}. For a compacted trie, if we extend in the natural way the
values of $E$ and $I$ the formula is no longer valid. In this note we provide a suitable generalization of the formula,
using the definition of \emph{extent} of a node (which collapses to the
definition of path when all compacted paths are empty). We show that $E=I+\TS$,
where $E$ is the sum of the lengths of external extents,
$I$ is the sum of the lengths of internal extents, and $\TS$
is the \emph{trie measure}, which approximates the number of bits that are
necessary to describe the trie. If all compacted paths are empty the trie measure
is $2n-2$, so our equation is a generalization of the classical result. 
We also provide a generalization to the case of non-binary tries.

\section{Definitions}

We work out our definitions from scratch closely following Knuth's, as the
notation that can be found in the literature is not always consistent.

\paragraph{Binary trees.} A \emph{binary tree} is either the empty binary
tree or a pair of binary trees (called the
\emph{left subtree} and the \emph{right
subtree})~\cite[page 312]{KnuACPI}.\footnote{We remark that the definition we
use (a slight abstraction on Knuth's) is the simplest and most correct from a
combinatorial viewpoint, but might sound unfamiliar. An alternative commonly
found description says that a binary tree is given by a node with a
left and a right subtree, either of which might be empty; the latter definition, however, does not account for the empty binary tree, which is essential in making the
left-child-right-sibling isomorphism with ordered forests work~(see
again~\cite[pages 334--335]{KnuACPI}).} 

A binary tree can be represented as a rooted tree\footnote{An acyclic connected
graph with a chosen node (the root). As observed by Knuth~\cite[page
312]{KnuACPI}, a tree (in the graph-theoretical sense) and a binary tree are two
completely different combinatorial objects.} in which nodes are either
\emph{internal} or \emph{external}. The empty binary tree is represented by a single external root node. Otherwise, a binary tree is represented by an internal root node connected to the representations of the left and right subtree by two edges labelled 0 and 1. Note that external nodes have no children, whereas internal nodes have always exactly two children.\footnote{We
remark that it is common to forget about external nodes altogether and consider
only internal nodes as ``true'' nodes of the binary tree. In this setting, there
are nodes with no children, nodes with a single child (left or right), and nodes
with two children. As noted by Knuth, handling external nodes explicitly makes
the structure ``more convenient to deal with''. In our case, external nodes are
essential in the very definition of $E$.}

\paragraph{Compacted binary tries.} A \emph{compacted binary trie} is either a
binary string, called a \emph{compacted path}, or a binary string endowed with a pair of binary tries
(called the \emph{left subtrie} and the \emph{right subtrie}). Equivalently,
a compacted binary trie can be seen as a labelling of the
nodes of a binary tree with compacted paths.

Given a nonempty prefix-free set of strings $S\subseteq
2^*$, the associated compacted binary trie is:
\begin{itemize}
  \item the only string in the set, if $|S|=1$;
  \item otherwise, let $p$ be the longest common prefix of the strings in $S$;
  then, the trie associated to $S$ is given by the string $p$ and by the pair
  of tries associated with the sets $\{\,x\in 2^*\mid pbx\in S\,\}$, for
  $b=0$, $1$.
\end{itemize}

A compacted binary trie can be represented as a rooted tree in which,
as in the case of binary trees, nodes are either \emph{internal} or
\emph{external}. A single string is represented by a single external root node
labelled by the string. Otherwise, a string and a pair of subtries are
represented by an internal root node labelled by the string, connected to the representations of the first and second subtrie by two
edges labelled 0 and 1 (see Figure~\ref{fig:names}).
From this representation, the set $S$ can be recovered by looking at the
labelled paths going from the root to \emph{external} nodes.

Given a node $\alpha$ of the trie (see again Figure~\ref{fig:names}):
\begin{itemize}
	\item the \emph{extent} of $\alpha$ is the longest
	common prefix of the strings represented by the external nodes 
	that are descendants of $\alpha$;
	\item the \emph{compacted path} of $\alpha$, denoted by $c_\alpha$, is the
	string labelling $\alpha$;
 	\item the \emph{name} of $\alpha$ is the extent of $\alpha$
 	deprived of its suffix $c_\alpha$.
\end{itemize}
We will use the name \emph{internal extent} (\emph{external extent}, resp.) for
the extent of an internal (external, resp.) node.

\begin{figure}\centering%
	\begin{tabular}{clccl}
    $s_0$ & 001001010\\
    $s_1$ & 00100110100100010\\
    $s_2$ & 001001101001001
    \end{tabular}
\caption{\label{fig:set}A toy example set $S$.}
\end{figure}

\begin{figure}\centering%
\includegraphics[scale=1]{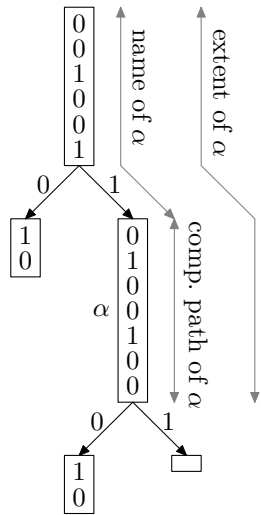}
\caption{\label{fig:names}The rooted-tree representation of the compacted trie
associated with the set $S$ of Figure~\ref{fig:set}, and the related names.
Arrows display the direction from the root to the external
nodes.}\end{figure}

\noindent\textbf{A data-aware measure.} Consider the compacted trie associated
with a nonempty set $S\subseteq 2^*$. We define the
\emph{trie measure} of $S$~\cite{GHSCDS} as
\[
\TS(S) 
= \sum_\alpha
(|c_\alpha| + 1) - 1=O(n\ell)
\]
where the summation ranges over all nodes of the trie, $n=|S|$ and $\ell$ is
the average length of the elements of $S$. Actually, $\TS(S)$ is the number of
edges of the standard (non-compacted) trie associated with $S$.

This measure is directly related to the number of bits
required to encode the compacted trie associated with $S$ explicitly: indeed,
to do this we just need to encode the trie structure (as a binary tree) and to write down in
preorder all the $c_\alpha$'s. Since there are $n$ external
nodes (hence $n-1$ internal nodes), writing a concatenation of the
$c_\alpha$'s requires $\TS(S)-2n+2$ bits; then we need
$\log{\TS(s) \choose 2n - 2 }$ additional bits to store the starting point of
each $c_\alpha$, whereas the trie structure needs just $2n-2$ bits (e.g., using
Jacobson's representation for binary trees~\cite{JacSSTG}). All in all, the space required to store the trie is
\[
	\TS(S) + \log{\TS(S) \choose 2n - 2 }.
\]
More precisely, the above number of bits is sufficient to write every trie with
$n$ external nodes and measure $\TS(S)$, and it is necessary for at least one
such trie (whichever representation is used)~\cite{FGGSCSCCO}.

\section{$E=I+\TS$}
We start by generalizing the internal path formula for binary trees to an
\emph{internal extent formula} for compacted binary tries:
\begin{theorem}
\label{thm:binary}
Let $S$ be a nonempty prefix-free set of $n$ binary strings with average length $\ell$,
and consider the compacted binary trie associated with $S$. Let $E$ be the sum
of the lengths of the external extents (equivalently: $E=n\ell$, the sum 
of the lengths of the strings in $S$), $I$ the sum of the lengths of the internal extents, and $\TS$ the trie measure of $S$.
Then,
\[
E=I+\TS.
\]
\end{theorem}
\begin{proof}
We prove the theorem by induction on $n$. The theorem is obviously true for 
$n=1$, as in this case $E=|c_\alpha|$, $I=0$ and
$\TS=|c_\alpha|+1-1=|c_\alpha|$. Consider now the case of a trie with root
$\alpha$ and subtries with their values $n_0$, $n_1$, $E_0$, $E_1$, $I_0$, $I_1$, $\TS_0$, and $\TS_1$. Then, using the definitions,
we have
\begin{align*}
E &= (E_0 + E_1) + (|c_\alpha|+1)(n_0+n_1) \\
I &= (I_0 + I_1) + (|c_\alpha|+1)(n_0-1+n_1-1) +  |c_\alpha|\\
  &= (I_0 + I_1) + (|c_\alpha|+1)(n_0+n_1-1) - 1\\
\TS &= (\TS_0+1) +( \TS_1+1) + (|c_\alpha|+1) -1 =\TS_0 + \TS_1 + |c_\alpha| +2
\\ n &= n_0 + n_1.
\end{align*}
Adding the equations $E_j=I_j + \TS_j$ for $j=0$, $1$ (which hold by
inductive hypothesis) we have
\[
E_0 + E_1 = I_0+ I_1 + \TS_0 + \TS_1.
\]
We add $(|c_\alpha|+1)(n_0+n_1)$ to both sides, getting 
\begin{eqnarray*}
E_0 + E_1 + (|c_\alpha|+1)(n_0+n_1)&=&  I_0+ I_1 +(|c_\alpha|+1)(n_0+n_1-1) + \TS_0 + \TS_1 +
|c_\alpha|+1\\
E_0 + E_1 + (|c_\alpha|+1)(n_0+n_1)&=& I + 1 + T - 1,
\end{eqnarray*}
which entails the thesis. 
\end{proof}
As noted in the introduction, when all compacted paths are empty $E$ is equal
to the external path length, $I$ is equal to the internal path length, and the
trie measure is exactly $\bigl(\sum_\alpha1\bigr)-1=2n-2$. Thus, the internal
extent formula is truly a generalization of the internal path formula.

\section{A simple application}

We were lead to the equation $E=I+\TS$ by the problem of bounding the average
length of an internal extent in terms of the average length of an external
extent, that is, in terms of $\ell$, the average length of the strings in $S$.
This bound can now be easily obtained:

\begin{corollary}
Let $|S|\geq2$ be a set of binary strings. With the notation of
Theorem~\ref{thm:binary},\[I/(n-1)\leq \ell - 3/2 + 1/n.\]
\end{corollary}
\begin{proof}
We just divide both members of the internal extent equation by $n$: 
\begin{align*}
\frac En &= \frac In + \frac Tn\\
 &= \frac I{n-1} + \frac In - \frac I{n-1} + \frac{2n - 2
+\sum_\alpha|c_\alpha|}n\\  &= \frac I{n-1} + \frac32 - \frac1n - \frac {I -
(n-1)(n/2 - 1 +\sum_\alpha|c_\alpha|)}{n(n-1)}\\ &\geq \frac I{n-1} +
\frac32-\frac1n.
\end{align*}
To see why the last bound is true, note that in a trie with $n-1$ internal
nodes the contribution to $I$ of the edges (i.e., excluding the compacted
paths) is at most $(n-1)(n-2)/2$ (the worst case is a linear trie). On the
other hand, the contribution of compacted paths to each internal path cannot be
more than $\sum_\alpha|c_\alpha|$, so the overall contribution cannot be more
than $(n-1)\sum_\alpha|c_\alpha|$. We conclude that
\[I\leq(n-1)\Bigl((n-2)/2+\sum_\alpha|c_\alpha|\Bigr).\qed\]
\end{proof}

Note that the bound is essentially tight, as in a linear trie with empty
compacted paths $E=n(n+1)/2-1$ and $I=(n-2)(n-1)/2$, so $E/n-I/(n-1)=3/2 -1 /n$.

\section{A generalization to non-binary tries}

Given an alphabet $\Sigma$, a \emph{compacted
trie over $\Sigma$} is defined as follows: it is either a
single string $x \in \Sigma^*$, or a string $x \in \Sigma^*$ together with a
subset $X\subseteq \Sigma$ with $|X|>1$ endowed with a function $\zeta$ that
assigns a compacted trie over $\Sigma$ to each element of $X$.

Given a nonempty prefix-free set of strings $S\subseteq \Sigma^*$, the
associated compacted trie over $\Sigma$ is:
\begin{itemize}
  \item the only string in the set, if $|S|=1$;
  \item otherwise, let $p$ be the longest common prefix of the strings in $S$;
  then, the trie associated with $S$ is given by $p$, the set $X\subseteq
  \Sigma$ of all $a \in \Sigma$ such that $pa$ is the prefix of some string in
  $S$, and by the function $\zeta$ mapping $a$ to the compacted
  trie associated with the set $\{\,x \in \Sigma^* \mid pax \in S\,\}$.
\end{itemize}

Similarly to what happens for compacted binary tries, a compacted trie over
$\Sigma$ can be represented as a rooted tree where each node is labelled by a
(possibly empty) string over $\Sigma$ and internal nodes have at most $|\Sigma|$
(but not less than two) children, each associated with a distinct symbol of
$\Sigma$. The notation of Figure~\ref{fig:names} carries on easily, and the
definition of trie measure is extended in the natural way.


We now want to generalize the internal extent formula (Theorem~\ref{thm:binary})
to nonbinary tries.
\begin{theorem}
Let $S$ be a nonempty prefix-free set of $n$ strings over an alphabet with
$\sigma$ symbols, and consider the compacted trie associated with $S$. 
For each $d=0,\dots,\sigma$, let $Y(d)$ be the sum of the lengths of the
extents of nodes with $d$ children, $n(d)$ be the number of
such nodes, and $T$ be the trie measure. Then,
\[
	E=\sum_{d=2}^\sigma (d-1)Y(d) + T.
\]
\end{theorem}
\begin{proof}
By induction on the number of nodes. This is true for a one-node trie; for the
induction step, suppose that the root of a trie has a compacted path of length $c$, and $h$ subtries
($2\leq h\leq\sigma$); for the $i$-th subtrie, by induction hypothesis,
since $E=Y(0)$ we have
\begin{equation}
\label{eqn:ihgen}
	Y_i(0)=\sum_{d=2}^\sigma (d-1)Y_i(d) + T_i.
\end{equation}
 
Observe that, for every $d=0,2,3,\dots,\sigma$,
\[
	Y(d)=\sum_{i=1}^h Y_i(d) + (c+1) \left( [d=h]+\sum_{i=1}^h n_i(d) \right) - [d=h]
\] 
where we used Iverson's notation.\footnote{For a given Boolean predicate $\phi$,
we let $[\phi]$ be 0 if $\phi$ is false, 1 if $\phi$ is true~\cite{KnuTNN}.} Moreover
\[
	n(d)=[d=h] + \sum_{i=1}^h n_i(d)
\]
so
\[
	Y(d)=\sum_{i=1}^h Y_i(d) + (c+1)n(d) - [d=h].
\]
Further
\[
	T=\sum_{i=1}^h T_i+h+c.
\]
Summing (\ref{eqn:ihgen}) memberwise, we obtain
\[
	\sum_{i=1}^h Y_i(0) = \sum_{d=2}^\sigma (d-1)\left(\sum_{i=1}^h Y_i(d)\right)
	+ \sum_{i=1}^h T_i
\]
that is equivalent to
\[
	Y(0) - (c+1) n(0) + [0=h] = \sum_{d=2}^\sigma
	(d-1)\left(Y(d)-(c+1)n(d)+[d=h]\right)+T-h-c,
\]
hence
\[
	Y(0)=\sum_{d=2}^\sigma (d-1)Y(d) - (c+1) \left(\sum_{d=2}^\sigma (d-1) n(d) -
	n(0)\right) + h - 1 +T-h -c.
\]
Since $\sum_{d=2}^\sigma (d-1) n(d)=n(0)-1$, we have
\[
	Y(0)=\sum_{d=2}^\sigma (d-1)Y(d) + T.\qed
\]
\end{proof}

\section{Acknowledgments}

We would like to thank the anonymous referee for spotting subtle inconsistencies
in the first version of this paper.

\bibliography{biblio}

\end{document}